\documentclass{article}
\usepackage{amsmath}
\usepackage{amssymb}
\usepackage{epsfig}
\usepackage{cite}

\newcommand{\be}{\begin{equation}}
\newcommand{\ee}{\end{equation}}
\newcommand{\bea}{\begin{eqnarray}}
\newcommand{\eea}{\end{eqnarray}}
\newcommand{\nn}{\nonumber}

\begin{document}
\begin{titlepage}

\begin{flushright}
Edinburgh 2006/15\\
August 2006
\end{flushright}
\vspace{1.cm}

\begin{center}
\large\bf
{\LARGE\bf Multi-Higgs boson production in the \\ Standard Model and beyond
}\\[1cm]
\rm
{ T.~Binoth$^{a}$, S.~Karg$^{b}$, N.~Kauer$^{b}$ and R.~R\"uckl$^{b}$  }\\[1cm]

{\em $^{a}$School of Physics, The University of Edinburgh,\\ 
	   Edinburgh EH9 3JZ, Scotland UK} \\[.5cm]%

{\em $^{b}$Institut f\"ur Theoretische Physik und Astrophysik,\\
           Universit\"at W\"urzburg, D-97074 W\"urzburg,
	   Germany} \\[.5cm]
     
\end{center}
\normalsize


\vspace{1cm}

\begin{abstract}
We present a calculation of the loop-induced 
processes $gg\to HH$ and $gg\to HHH$, and investigate the observability
of multi-Higgs boson production at the CERN Large Hadron Collider (LHC)
in the Standard Model (SM) and beyond.
While the SM cross sections 
are too small to allow observation at the LHC, we demonstrate 
that physics beyond the SM can lead
to amplified, observable cross sections.
Furthermore, the applicability of the heavy top quark
approximation in two- and three-Higgs boson production is investigated.
We conclude that multi-Higgs boson production at the SuperLHC 
is an interesting probe of Higgs sectors beyond the SM and warrants further study.
\end{abstract}



\end{titlepage}

\section{Introduction}

The understanding of electroweak symmetry breaking is a major
goal of the physics programme at the upcoming Large Hadron Collider
(LHC) at CERN. In the Standard Model (SM), electroweak symmetry breaking 
is mediated by the Higgs mechanism, which predicts a fundamental 
scalar particle, the Higgs boson \cite{Higgs:1964ia}.  
At the LHC, the discovery  of the Higgs boson is feasible 
in the entire theory-compatible mass region 
\cite{Buscher:2005re,Djouadi:2005gi}.
Theoretically, a heavy Higgs boson with mass of order 1 TeV in the vicinity of the 
non-perturbative regime is still feasible \cite{Binoth:1998tg,Ghinculov:1998km}. 
Electroweak precision measurements, however, favour a relatively light SM Higgs boson
not too far above the direct LEP exclusion limit of 114.4 GeV \cite{Barate:2003sz}.

In order to establish the Higgs mechanism and confirm the Higgs sector of the SM,
the discovery of a Higgs-like boson is not sufficient. 
In addition, the predicted couplings of fermions and gauge bosons to the Higgs boson, 
as well as the trilinear and quartic Higgs self couplings need to be confirmed experimentally.
While Higgs-fermion and Higgs-gauge boson couplings
are measurable with accuracies of 10--40\% at the LHC \cite{Duhrssen:2004cv} and in
many channels considerably better at the International Linear Collider 
(ILC) \cite{Weiglein:2004hn}, the measurement of the trilinear 
and quartic Higgs self couplings, which are probed in double and triple Higgs boson 
production, respectively, are more challenging.
For low Higgs masses not too much above the LEP limit ($m_H\lesssim140$ GeV),
the largest rates are obtained with the decay channel $H\to b\bar b$ dominating in this mass regime.
Unfortunately, at the LHC this search channel is not viable, because of an overwhelming QCD background.  The ILC, however, 
would allow the measurement of the  trilinear Higgs self coupling to a precision 
of 20-30\% in the low Higgs mass regime \cite{Weiglein:2004hn,Djouadi:1999gv,Battaglia:2001nn,Gutierrez-Rodriguez:2006qk}.
We note that in this regime a photon collider promises an even better  
determination of the trilinear Higgs coupling \cite{Jikia:1992mt,Boudjema:1995cb,Belusevic:2004pz}.  For higher Higgs masses, the ILC production cross section decreases due to the reduced phase space.  However, in this mass region, the vector boson pair decay channels open up and allow for leptonic signatures that can be separated
from the backgrounds at the LHC 
\cite{Weiglein:2004hn,Djouadi:1999rc,Baur:2002qd,Baur:2002rb,Baur:2003gp,Baur:2003gpx}.       

The dominant production mechanism for Higgs boson pairs at the LHC is gluon fusion.
We note that Higgs production in hadronic collisions can also proceed through bottom 
quark fusion, $b\bar{b} \to nH$, but in the SM the corresponding LO as well as NLO
\cite{Dawson:2006dm} cross sections are negligible. In the MSSM, however, enhanced Yukawa
couplings can lead to comparable cross sections for gluon and bottom quark fusion \cite{Jin:2005gw}. 

The gluon fusion loop amplitude was first presented in \cite{Glover:1987nx}.
For neutral Higgs boson pairs in the Minimal Supersymmetric Standard Model (MSSM) the
top/bottom loop contribution was evaluated in \cite{Plehn:1996wb,Krause:1997rc}.
Charged Higgs boson pairs including squark effects were studied in \cite{Brein:1999sy}.
Although the SM cross section for triple Higgs boson production at the LHC can be expected 
to be small
\cite{Djouadi:1999gv,Battaglia:2001nn}, this expectation has to be verified
through explicit calculation.
Only recently, a full calculation of the process $gg\to HHH$ appeared in the literature 
\cite{Plehn:2005nk} and confirmed that SM cross sections are indeed too small to be 
observable at the LHC.  However, it has to be stressed that multi-Higgs boson production
rates are very sensitive to physics beyond the Standard Model (BSM) and 
should be scrutinised carefully in high energy experiments, as they might point to 
new physics at high energy scales that are not directly accessible at the given collider \cite{Kanemura:2002vm,Kanemura:2004mg}. 
A non-standard heavy quark that
receives its mass via the Higgs
mechanism  does not decouple \cite{Appelquist:1974tg},
and therefore leads to a non-vanishing contribution in heavy quark loop-induced 
processes.  
Furthermore, contributions of higher dimensional operators 
might alter the SM cross section considerably \cite{Barger:2003rs}.
Also, in certain Little Higgs Models the Higgs pair production cross section is significantly different from the SM cross section \cite{Dib:2005re}.

In this paper, we present our calculation of double and triple Higgs boson production
via gluon fusion.  This  provides an independent check of the
recent calculation in \cite{Plehn:2005nk}, which employed 
different computational methods and tools. 
To go beyond the findings of \cite{Plehn:2005nk}, we study higher dimensional operator effects on production rates relative to the SM. The corresponding couplings are in principle only
restricted by unitarity constraints \cite{Lee:1977eg}.
We also analyse amplification effects in supersymmetric (SUSY) two-Higgs-doublet models (2HDMs).

The paper is organised as follows.  After briefly reviewing Higgs boson
properties relevant for our investigation in Section 2, we describe our
calculation in Section 3. In Section 4, SM and BSM results are presented and 
discussed.  Section 5 gives a brief summary.

\section{Higgs boson properties}

The SM Higgs mechanism provides fundamental mass terms for
massive vector bosons and fermions.  The coupling strength
of the Higgs boson is proportional to the mass (squared) of the 
interacting fermion (gauge boson).  The SM Higgs boson self interactions, induced by the scalar potential 
\bea
V = \frac{m_H^2}{2\, v^2} \Bigl( \Phi^\dagger \Phi - \frac{v^2}{2}\Bigr)^2 \ ,
\eea
are also proportional to the Higgs mass squared. 
In unitary gauge one has
\bea
V = \frac{m_H^2}{2} \, H^2 + \frac{\lambda_3}{3!} \, H^3 + \frac{\lambda_4}{4!}\, H^4 
\eea
with
\bea\label{lambda_rel}
   \lambda_4 = \lambda_3/v = \frac{3\, m_H^2}{v^2} \ .
\eea
Relation (\ref{lambda_rel}) between the trilinear and quartic Higgs self 
couplings 
is a genuine SM prediction.  To establish the SM Higgs mechanism, 
it has to be verified experimentally.  As already pointed out, the discovery of
the Higgs boson and the measurement of its couplings to fermions and gauge bosons
alone are not sufficient.

In general, the Higgs self couplings change in extensions of the SM.
By allowing for higher dimensional operators of the
type 
\bea
\sum\limits_{k=1}^{\infty}\frac{g_k}{\Lambda^{2k}} \left(\Phi^\dagger \Phi - \frac{v^2}{2}\right)^{2+k} \ ,
\eea
the constraint (\ref{lambda_rel}) is relaxed.  Magnitude and sign 
of $\lambda_3$ and $\lambda_4$ can be arbitrary up to constraints 
imposed by unitarity. In order to guarantee the stability of the 
vacuum, only the sign of the highest power of the Higgs field has to be
positive.
We note that the addition of singlet Higgs fields 
preserves relation (\ref{lambda_rel}), but
may lead to invisible Higgs boson decays and diluted
Higgs signals \cite{Binoth:1996au}. 

The ability to measure the Higgs self couplings depends on the
size of multi-Higgs boson cross sections. As will be discussed below,
SM rates are very small at the LHC  (see also \cite{Plehn:2005nk}). 
It is thus interesting to consider extensions of the SM that
allow for amplified  event rates.
For  Higgs pair production in gluon fusion this has been studied
in \cite{Plehn:1996wb} in the context of the MSSM. 
Two amplification sources have been identified. Firstly,
the top and bottom Yukawa couplings are altered due to the mixing 
of  the Higgs fields. In the MSSM one has, at tree level,
\bea\label{susy_yuk}
\lambda_{ht\bar t} &=& \frac{ m_t}{v} \frac{\cos \alpha}{\sin \beta}\ ,  \nn\\
\lambda_{hb\bar b} &=& -\frac{ m_b}{v} \frac{\sin \alpha}{\cos \beta}\ ,
\eea 
where
\bea
\tan \beta = \frac{v_2}{v_1} &\mbox{and}&
\alpha = \frac{1}{2} \arctan \left( \frac{M_A^2+M_Z^2}{M_A^2-M_Z^2} \tan 2\beta \right),  
             \quad  - \frac{\pi}{2} \leq \alpha \leq 0\ .
\eea
Here, $v_1$ ($v_2$) is the vacuum expectation value of the Higgs doublet with weak hypercharge $-\frac{1}{2}$ ($+\frac{1}{2}$). 
For sufficiently large $\tan \beta$, the bottom-loop contribution to the
cross section becomes sizable in comparison to the top-loop contribution, leading to a
larger production rate. Secondly, internal Higgs propagators can become resonant
thereby enhancing the production rate.
Both effects generally play a role in 2HDMs, as will be discussed and quantified
for triple Higgs boson production below.

We will now briefly review the features of 2HDMs that are important for our purposes.
The general potential of the 2HDM is given by \cite{Gunion:1989we,Haber:1993an}
\begin{eqnarray}\label{dimpara}
	V(\Phi_1,\Phi_2)& = &m_{11}^2\Phi_1^\dagger\Phi_1+m_{22}^2\Phi_2^\dagger\Phi_2-
        (m_{12}^2\Phi_1^\dagger\Phi_2+ {\rm h.c.})+ 
\nonumber\\
&&       + \frac{\lambda_1}{2}(\Phi_1^\dagger\Phi_1)^2+
        \frac{\lambda_2}{2}(\Phi_2^\dagger\Phi_2)^2+ 
\nonumber\\ 
&&	+
        \lambda_3(\Phi_1^\dagger\Phi_1)(\Phi_2^\dagger\Phi_2)
	+\lambda_4(\Phi_1^\dagger\Phi_2)(\Phi_2^\dagger\Phi_1)+
\nonumber \\
&&       
	+\left\{\frac{\lambda_5}{2}(\Phi_1^\dagger\Phi_2)^2
	+[\lambda_6(\Phi_1^\dagger\Phi_1)+\lambda_7(\Phi_2^\dagger\Phi_2)]
        (\Phi_1^\dagger\Phi_2)+ {\rm h.c.}\right\}\ ,
\end{eqnarray} 
with the complex Higgs-doublet fields acquiring the vacuum expectation values $v_1$ and $v_2$. After
diagonalising the mass matrix one obtains the physical Higgs fields $h,H,A,H^{\pm}$ and the Goldstone bosons 
$G^{\pm},G$. We are interested in the  Higgs self couplings, which  can be written in terms of the dimensionless
parameters $\lambda_i$, $i=1,\ldots,7$, appearing in (\ref{dimpara}). 
For the quartic couplings, one has
\begin{eqnarray}
\lambda_{hhhh} &=& 3\cos^4 \alpha \lambda_2 + 3\sin^4 \alpha \lambda_1 + 6 \cos^2 \alpha \sin^2
\alpha  \;(\lambda_3+\lambda_4+\lambda_5) 
\nn\\&&
- 12 \cos^3 \alpha \sin \alpha \lambda_7 - 
 12 \sin^3 \alpha \cos \alpha \lambda_6 \ ,\nn \\
 \lambda_{Hhhh} &=& -3 \cos \alpha \sin^3 \alpha \lambda_1 + 3 \cos^3 \alpha \sin \alpha \lambda_2
   - \frac{3}{2} \cos 2\alpha \sin 2\alpha \;(\lambda_3+\lambda_4+\lambda_5) 
\nn\\&&
   +3(3\cos^2 \alpha \sin^2 \alpha - \sin^4\alpha)\lambda_6 + 3(\cos^4 \alpha - 3 \cos^2 \alpha \sin^2 \alpha)\lambda_7 \ ,\nn 
\end{eqnarray}    
to be multiplied by $M_Z^2/v^2$. 
In the MSSM, the Higgs sector is constrained such 
 that only two of the seven input parameters are free. Choosing $\tan \beta$ and $M_A=\lambda_6
v^2$ as basic input, one recovers the  MSSM values
\begin{eqnarray}
\lambda_{hhhh} &=& 3 \cos^2 2\alpha \ ,\nn \\
\lambda_{Hhhh} &=& 3  \cos 2\alpha \sin 2 \alpha\ .
\end{eqnarray}  
As is well known, radiative corrections in the Higgs sector are large. 
The largest ubiquitous correction is given by $ 3G_F m_t^4/(\sqrt2 \pi^2 \sin^2 \beta)\cdot
\ln(m^2_{\tilde t}/m^2_t)$.
In our  numerical analysis for the MSSM all one-loop and leading two-loop corrections to Higgs masses and couplings
are included \cite{Carena:1995bx,HDECAY}.
For a discussion about reconstructing the Higgs potential
in the SUSY case, see \cite{Boudjema:2001ii}.
 
\section{Calculation}

From a computational point of view loop amplitudes with five or more
external legs are challenging due to their combinatorial complexity.
We only sketch here our method, which leads to a 
numerically stable algebraic representation of the amplitude.  
More details can be found in \cite{Binoth:1999sp,Binoth:2003xk,Binoth:2005ua,Binoth:2005ff,Binoth:2006rc,Binoth:2006mf}.

\subsection{Structure of the amplitude}

The production processes $pp\to hh$ and $pp\to hhh$ proceed at the parton level via gluon fusion in combination with a quark loop. 
Here $h$ stands for the   Higgs boson of the SM or the light CP-even Higgs boson of 
the MSSM. Squarks in the loop have been neglected.
Their contributions vanish for large squark masses, due to their decoupling property.
   In the SM, only the top
quark loop leads to a non-negligible cross section at the LHC. 
For triple Higgs boson production,
\bea \label{amp}
g(p_1,\lambda_1) + g(p_2,\lambda_2) \to h(p_3) + h(p_4) + h(p_5) \ ,
\eea
with $p_i$ defining the 4-momenta and $\lambda_{1,2}$ specifying the gluon helicities,
the Feynman diagrams are classified in Fig.~\ref{graphs}.
On the diagrammatic level 3-, 4- and 5-point topologies can be distinguished.
\unitlength=1mm
\begin{figure}[ht]
\begin{picture}(150,75)
\put(5,65){$P$:}
\put( 15,38){\includegraphics[width=4.5cm]{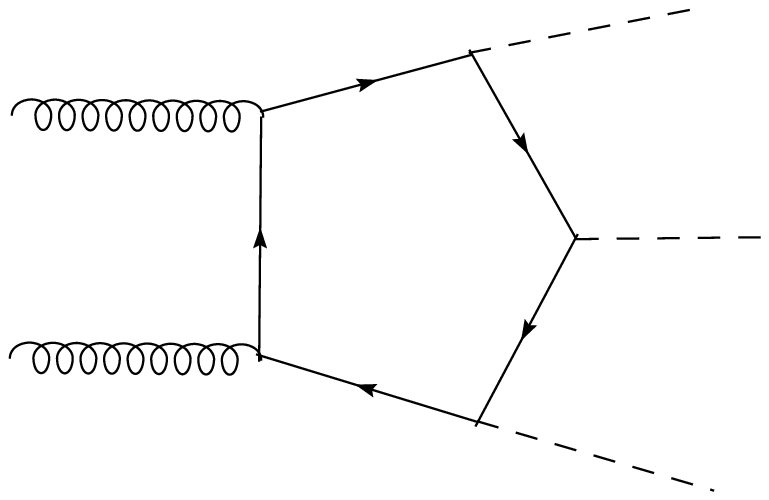}}
\put(80,65){$B$:}
\put( 90,38){\includegraphics[width=4.5cm]{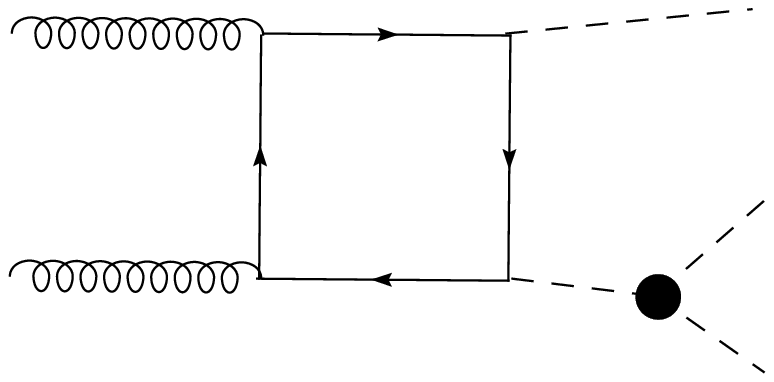}}
\put(5,30){$T_1$:}
\put( 15,7 ){\includegraphics[width=4.5cm]{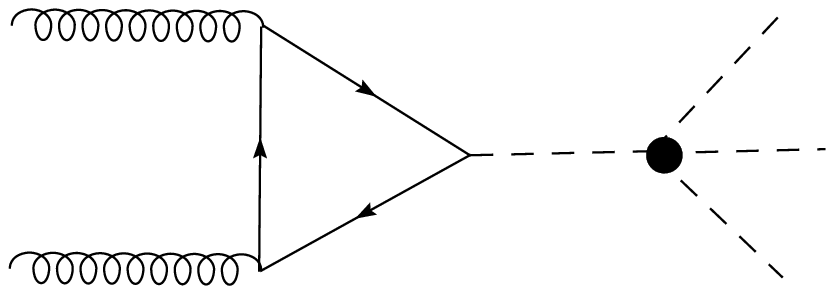}}
\put(80,30){$T_2$:}
\put( 90,7 ){\includegraphics[width=4.5cm]{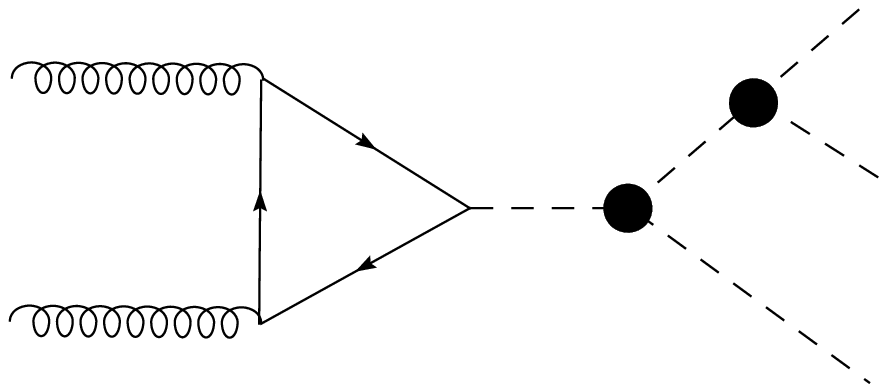}}
\end{picture}	
\caption{\label{graphs}Illustration of different topologies to the process $gg\to hhh$.}
\end{figure}
Each topology involves a different combination of coupling constants. 
Here, we only discuss the structure of the SM amplitude. In  2HDMs one has
additional tree structures from the heavy Higgs boson $H$ attached to
the quark loop.
The pentagon topology $P$ contains no Higgs self coupling. The box topologies $B$
are proportional to $\lambda_{3}$, whereas the triangle topologies $T_1$ and $T_2$ are
proportional to $\lambda_{4}$  and $\lambda_{3}^2$, respectively.
Subsequently, the  amplitude can be expressed as 
\bea
\Gamma(gg\to hhh) = \delta_{ab} \, T_R \, \frac{\alpha_s}{4\pi}\,\,
\varepsilon_{1,\mu}\varepsilon_{2,\nu}  \,\,{\cal M}^{\mu \nu}  \ ,
\eea
\bea
{\cal M}^{\mu \nu} =  \lambda_{tth}^3 \,{\cal M}^{\mu \nu}_{P}  
                          + \lambda_{3}\,\lambda_{tth}^2\,  {\cal M}^{\mu \nu}_{B}  
			  + \lambda_{4} \,\lambda_{tth} \,{\cal M}^{\mu \nu}_{T,1} 
			  + \lambda_{3}^2 \,\lambda_{tth} \,{\cal M}^{\mu \nu}_{T,2} \ .
\eea
 
The scattering tensor ${\cal M}^{\mu \nu}$ can be decomposed
in terms of metric tensors and  external momenta and, using momentum conservation,
be expressed as
\bea 
{\cal M}^{\mu \nu} &=& \hat A \, g^{\mu \nu} + \sum\limits_{j,l=1,4} \hat B_{jl} \,\, p_j^{\mu}p_l^{\nu}\ .
\eea
By solving the Ward identities ${\cal M}^{\varepsilon_1 p_2}=0$, ${\cal M}^{p_1 \varepsilon_2}=0$, 
or equivalently using axial gauge conditions (e.g. $\varepsilon_1 \cdot p_2$, 
$\varepsilon_2 \cdot p_1$), one can achieve a manifestly gauge invariant representation
of the amplitude. Introducing the Abelian part of the gluon field strength tensor
${\cal F}_j^{\mu\nu}= \varepsilon_j^\mu p_j^\nu -  p_j^\mu\varepsilon_j^\nu$ it reads
\bea 
{\cal M}^{\varepsilon_1\varepsilon_2} &=&  A \, \textrm{tr}( {\cal F}_1 {\cal F}_2 )  
       + \sum\limits_{j,l=3,4} B_{jl} \,\,p_2\cdot {\cal F}_1 \cdot p_j\;  p_1\cdot {\cal F}_2 \cdot p_l \ .
\eea
The amplitude coefficients $A$, $B_{jl}$ are equal to $\hat A$, $\hat B_{jl}$,
up to trivial factors.
Bose symmetry of the gluons and Higgs bosons leads to additional relations between them.
After determining all amplitude coefficients, verifying the Ward identities and Bose symmetry
served as a powerful check of our calculation. 

It is useful to decompose the amplitude further into helicity components.
Due to parity invariance only two helicity amplitudes have to be known:
\bea 
{\cal M}^{--} &=& {\cal M}^{++} \ ,\nonumber\\ 
{\cal M}^{-+} &=& {\cal M}^{+-} \ .
\eea
Applying spinor helicity methods \cite{Xu:1986xb,Dixon:1995xx},
the polarisation vectors for $\pm$ helicities are given by
\bea
\varepsilon_{1}^{+\;\mu} \varepsilon_{2}^{+\;\nu}&=& - \frac{[21]}{\langle
12\rangle} \frac{\textrm{tr}^-(1\nu 2\mu)}{2\,s_{12}} 
\ ,\nonumber\\
 \varepsilon_1^{+\;\mu} \varepsilon_2^{-\;\nu} &=&  \frac{ \langle 2^-|\mu| 1^- \rangle }{\sqrt{2} \langle 21 \rangle }
                                                    \frac{ \langle 2^-|\nu| 1^- \rangle }{\sqrt{2} [12]}\ ,
\eea
with $\textrm{tr}^-(1\nu\dots)\equiv [\textrm{tr}(p_1\!\!\!\!\!/ \gamma^\nu\dots) 
- \textrm{tr}( \gamma_5 p_1\!\!\!\!\!/\gamma^\nu\dots )]/2$ and the spinor inner products $\langle ij\rangle\equiv\langle p_i^-|p_j^+\rangle,\ [ij]\equiv\langle p_i^+|p_j^-\rangle$, where $|p_i^\pm\rangle$ is the Weyl spinor for a massless particle with momentum $p_i$.\\
This implies
\bea
{\cal M}^{++} = \frac{[21]}{\langle 12 \rangle} \Bigl(  
   A 
- \frac{\textrm{tr}^-(1323)}{ 2 s_{12} } \, B_{33}
- \frac{\textrm{tr}^-(1423)}{ 2 s_{12} } \, B_{34} \nn \\ \qquad
- \frac{\textrm{tr}^-(1324)}{ 2 s_{12} } \, B_{43}
- \frac{\textrm{tr}^-(1424)}{ 2 s_{12} } \, B_{44}
        \Bigr)
\eea
and
\bea
{\cal M}^{+-} = \frac{\langle 2^-|3|1^- \rangle}{\langle 1^-|3|2^- \rangle} \Bigl(  
 \;\; \frac{\textrm{tr}^-(1323)}{ 2 s_{12} } \, B_{33}
+ \frac{\textrm{tr}^-(1324)}{ 2 s_{12} } \, B_{34}\Bigr) \nn \\
+ \frac{\langle 2^-|4|1^- \rangle}{\langle 1^-|4|2^- \rangle} \Bigl(  
\;\; \frac{\textrm{tr}^-(1423)}{ 2 s_{12} } \, B_{43}
+ \frac{\textrm{tr}^-(1424)}{ 2 s_{12} } \, B_{44}
        \Bigr) \ .
\eea 
Contrary to the $++$ case, it is not possible to factor out
a global spinorial phase in the $+-$ case without introducing denominators 
that in general aggravate numerical problems.

\subsection{Evaluation of the amplitude coefficients}

Our goal was the analytical reduction of all diagrams, to allow for
algebraic cancellations of numerically dangerous denominators in the amplitude.
These denominators are so-called Gram determinants which are induced by 
reduction algorithms of Lorentz tensor integrals.
After generating all diagrams, using the QGRAF \cite{Nogueira:1991ex} program,
we used FORM 3.1 \cite{Vermaseren:2000nd} to perform the gamma matrix algebra
and to project the diagrams on the helicity components and amplitude coefficients.
Further, by applying the reduction algorithms for scalar and tensor integrals
described in \cite{Binoth:2005ff,Binoth:2006rc}, we expressed 
all amplitude coefficients
as a linear combination of  scalar integrals. As scalar integral basis we chose
2-, 3- and 4-point functions $(s_{ij}=(p_i+p_j)^2)$:
\bea 
&&I_2^{d=n}(s_{ij},m_q^2,m_q^2)\ , \nonumber\\
&&I_3^{d=4}(s_{ij},s_{kl},s_{pr},m_q^2,m_q^2,m_q^2)\ ,\nonumber\\ 
&&I_4^{d=6}(s_{ij},s_{kl},s_{pr},m_q^2,m_q^2,m_q^2,m_q^2)\ ,\nonumber
\eea
which were evaluated using
LoopTools-2.2 \cite{Hahn:2000kx}. The spurious UV pole of the 2-point integral cancels when adding all diagrams. The full amplitude is composed
out of 12, 24 and 31 different 2-, 3-, and 4-point functions. 
The complexity of the expressions is induced by the number of  independent scales,
which is seven here. One may chose $s_{12}$, 
$s_{23}$, $s_{34}$, $s_{45}$, $s_{15}$, $m_q^2$, $m_h^2$.
The coefficient of each function was  exported to MAPLE
to apply simplification algorithms. 
Schematically,
\bea
{\cal M}^{\lambda_1 \lambda_2} = \sum\limits_{k} \textrm{simplify}[C^{\lambda_1 \lambda_2}_k] 
  \: I_k \quad, \quad I_k \in \{I_2^n,I_3^4,I_4^{6}\} \ .
\eea
In this way, we could achieve expressions with
a simple denominator structure allowing for a stable numerical evaluation. 
In the equal helicity case, $\lambda_1=\lambda_2$, all Gram determinants cancel. In the
opposite helicity case, $\lambda_1=-\lambda_2$, one Gram determinant survives.
The simplified expressions were then exported to Fortran code.
Each of these steps was completely automatized. 

\subsection{Numerical implementation}
In order to compute numerical results for hadron colliders, the differential partonic cross section has to be convoluted with parton distribution
functions (PDFs) and integrated over the $2\to 3$-particle phase space.
We employed the gluon density of the MRST2002nlo PDF set \cite{Martin:2002aw}, as implemented in LHAPDF \cite{LHAPDF}, which also provides the strong coupling constant as function of the renormalisation scale.
In the MSSM case, where the 
heavy CP-even Higgs boson $H$ can be resonant, we used multi-channel Monte Carlo (MC)
integration techniques \cite{Berends:1994pv,Kleiss:1994qy} with phase space mappings based on \cite{Kauer:2001sp,Kauer:2002sn} and the adaptive MC integration package BASES \cite{BASES}.
The relevant quartic Higgs couplings were implemented in the program HDECAY \cite{HDECAY}, which incorporates 
the routine FeynHiggsFast \cite{FeynHiggs}, in order to evaluate the radiatively corrected Yukawa- and 
Higgs couplings and also the Higgs widths, as discussed above.

\section{Results}

In this section we present and discuss the LHC cross sections for 2- and 3-Higgs boson production.  We use the following parameters throughout:
\bea
\alpha_s(M_Z) &=& 0.120 \nn \\
\alpha(0)  &=& 1/137.036 \nn \\
m_t  &=& 178 \, \textrm{ GeV}\nn \\
m_b  &=& 4.7 \,\textrm{ GeV}\nn \\
m_W  &=& 80.41 \,\textrm{ GeV}\nn \\
m_Z  &=& 91.1875\, \textrm{ GeV}\nn \\
\eea  

For 2-Higgs (3-Higgs) boson production, the factorisation scale $\mu_F$ and renormalisation scale $\mu_R$ were set to $\mu_F = \mu_R = 2 m_H$ $(3 m_H)$.
The strong coupling constant $\alpha_s$ was taken at $\mu_R$, but for the
fine structure constant we used $\alpha(0)$.
All results have been calculated with a MC error of 0.5\% or less.

In the following subsections we present SM  cross sections
for 2- and 3-Higgs boson production and describe how BSM scenarios allow for observable enhancements
of the SM rates.

\subsection{Multi-Higgs boson production in the SM}

We begin with the gluon-fusion production cross section 
for 2 and 3 Higgs bosons. Note that in both cases large next-to-leading 
order corrections 
are expected,  leading to $K$-factors as large as 2 like in the case of single Higgs boson production \cite{Dawson:1990zj,Djouadi:1991tk}, since the infrared structure
of these processes is identical (with a large contribution 
from soft gluon effects).

\begin{figure}[ht]
\unitlength=1mm
\begin{picture}(180,70)
\put(20,0){\includegraphics[width=11.cm]{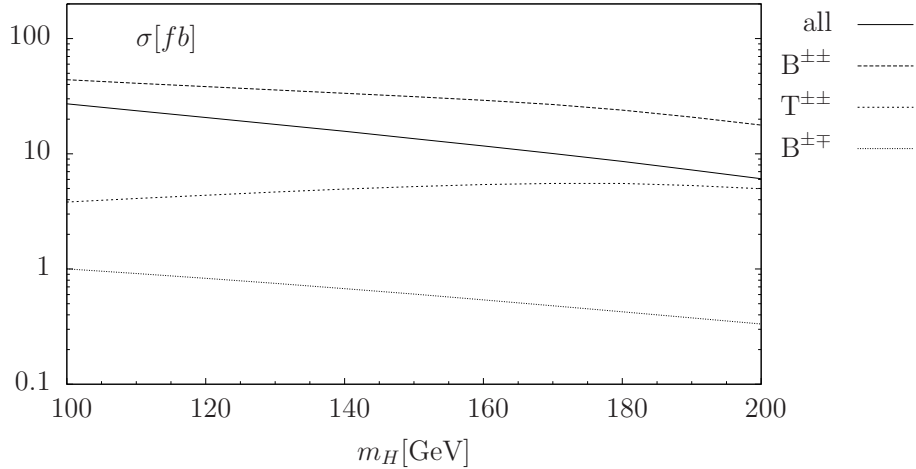}} 
\end{picture}
\caption{Total Higgs pair production cross section vs.~$m_H$ at the LHC, 
as well as the individual cross sections for box (B) and triangle (T) topologies with
equal ($++/--$) and opposite ($+-/-+$) helicity components. }
\label{fig:gghh}
\end{figure}

In Fig.~\ref{fig:gghh}, we display the Higgs pair production cross section 
vs.~$m_H$ at $\sqrt{s}$ = 14 TeV. 
Our results show good agreement with \cite{Plehn:1996wb}
when PDF, scale and parameter uncertainties are taken into account.
The total cross section falls from about 30 to 6 fb in the Higgs mass range
from 100 to 200 GeV.  
In addition to the total cross section, 
the equal  $++/--$ and opposite $+-/-+$ helicity components 
of the cross section are also shown.
The triangle topologies only allow for a $L=S=0$ interaction, 
i.e. the $+-/-+$ helicity component is zero.
Overall, the opposite helicity component is more than an order of magnitude suppressed.
Furthermore, a destructive interference effect is visible between the 
box and triangle topologies. 
\begin{figure}[ht]
\unitlength=1mm
\begin{picture}(180,70)
\put(20,0){\includegraphics[width=11.cm]{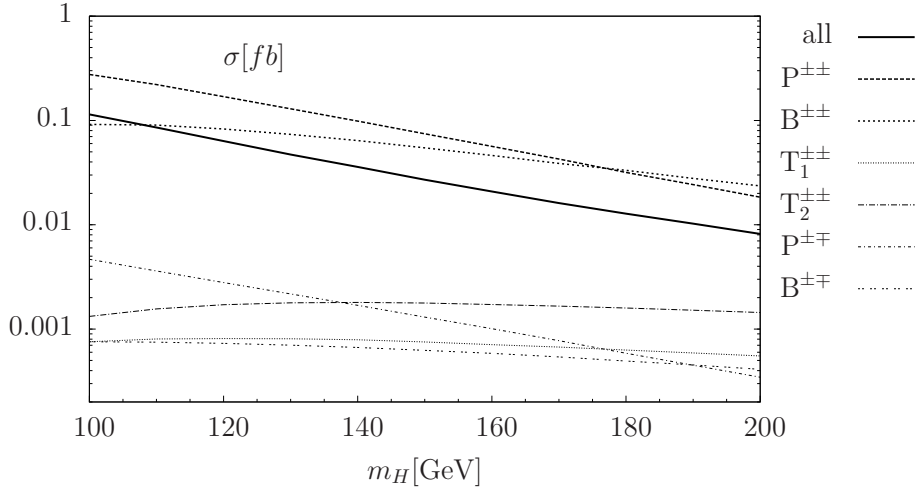}} 
\end{picture}
\caption{The total  3-Higgs boson production cross section vs.~$m_H$ at the LHC.
Equal and opposite helicity components of contributions 
from various topologies are shown.}
\label{fig:gghhh}
\end{figure}

In Fig.~\ref{fig:gghhh}, the total cross section for triple Higgs boson production at the LHC is
plotted vs.~$m_H$. Our results agree with the recent
calculation of \cite{Plehn:2005nk} within MC errors.
Again one finds that the opposite helicity 
components of the cross section are more than an order of magnitude
suppressed. In the figure, the pentagon, box,
and triangle contributions are also shown separately. 
The latter, being proportional to $\lambda_3^2$ 
and $\lambda_4$, are suppressed relative to the box and pentagon 
topologies. Due to interference effects 
the contribution of the quartic Higgs self coupling to the total cross section
is not negligible: it varies between +1\% ($m_H=100$ GeV) and --57.5\% ($m_H=200$ GeV).
The destructive interference pattern between triangle, box
and respectively box and pentagon contributions is well-known.
It can be understood from from the fact that the effective 
two-gluon $n$-Higgs boson operators contain a factor $(-1)^n$ \cite{Glover:1987nx}.
As the self couplings increase with increasing Higgs mass, the box and triangle 
topologies become more and more important relative to the pentagon contribution.

In Table \ref{Tab:Numbers} we give predictions for different 
values of the  Higgs mass for the LHC and a 200 TeV Very Large Hadron Collider (VLHC).     
\begin{table}
\vspace*{0.8cm}
\centering
\begin{tabular}{ccccc}
   \hline
   \hline
  & $m_H\, [\textrm{GeV}]$   & $120$ & $150$ & $180$ \\[0.5mm]
    \hline
 LHC& $\sigma \,[\textrm{fb}]$  & $0.0623  $ & $0.0267$ & $0.0126$ \\
 VLHC& $\sigma \,[\textrm{fb}]$ & $ 9.55  $ & $4.89$ & $ 2.98$ \\
  \hline
  \hline
\end{tabular}
\caption{Typical cross sections for triple Higgs boson production at the LHC and a 200 TeV VLHC.}
\label{Tab:Numbers}
\end{table}
We also note that a change in the top mass best fit value from $m_t=178$ GeV to $172.5$ 
GeV leads to a 15\% decrease of the cross section.

In Table ~\ref{Tab:VLHCNumbers}, we compare the relative importance of the different topologies
and helicities at the VLHC.
\begin{table}
\vspace*{0.8cm}
\centering
\begin{tabular}{cccccccc}
   \hline
   \hline
 $\sigma \,[\textrm{fb}]$  &  all & $P^{\pm\pm}$ & $B^{\pm\pm}$ & 
  $T_1^{\pm\pm}$& $T_2^{\pm\pm}$ & $P^{\pm\mp}$ & $B^{\pm\mp}$\\[0.5mm]
    \hline
 $m_H=120$ GeV& 9.55  & 21.82 & 10.04 & 0.111 & 0.189 & 0.589 & 0.169\\
 $m_H=200$ GeV& 1.93  &3.97  &4.76  &0.129  &0.262  &0.125 & 0.163\\
  \hline
  \hline
\end{tabular}
\caption{Contributions of different topologies to triple Higgs boson production at a 200 TeV VLHC.}
\label{Tab:VLHCNumbers}
\end{table}

\begin{figure}[ht]
\unitlength=1mm
\begin{picture}(150,50)
\put(0, 0){\includegraphics[width=7.cm]{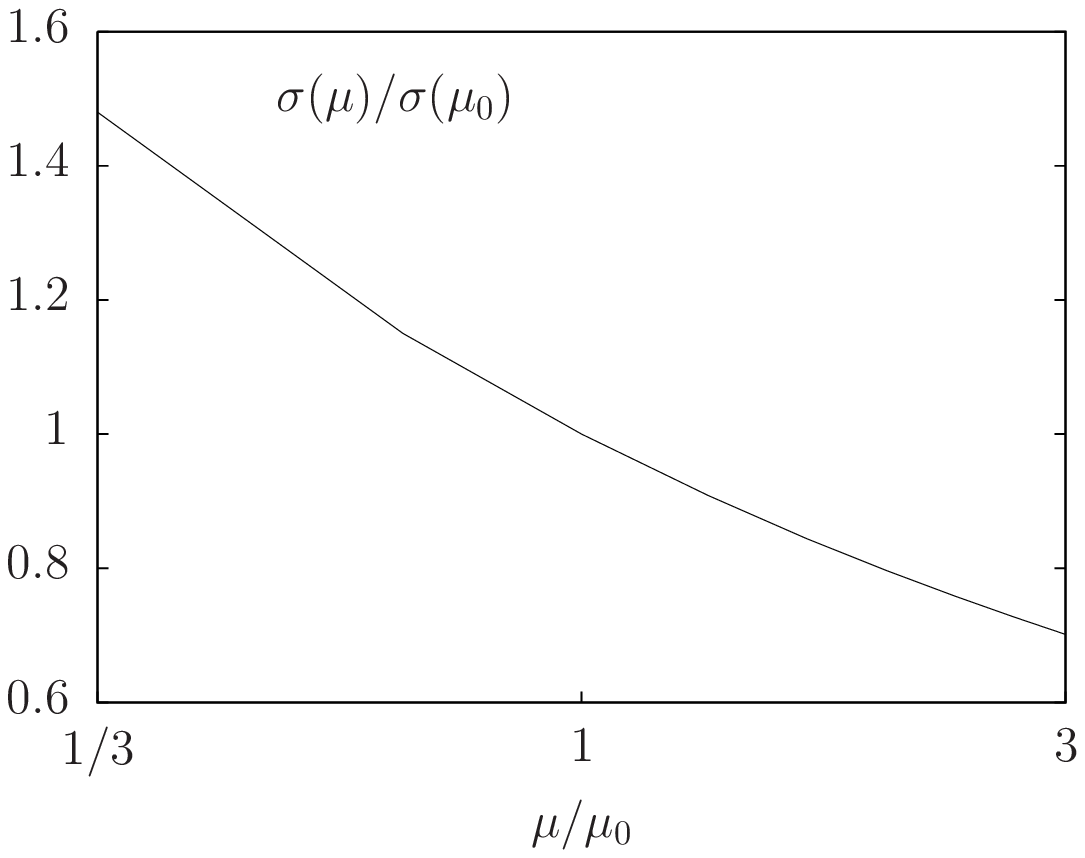}} 
\put(75,0){\includegraphics[width=7.cm]{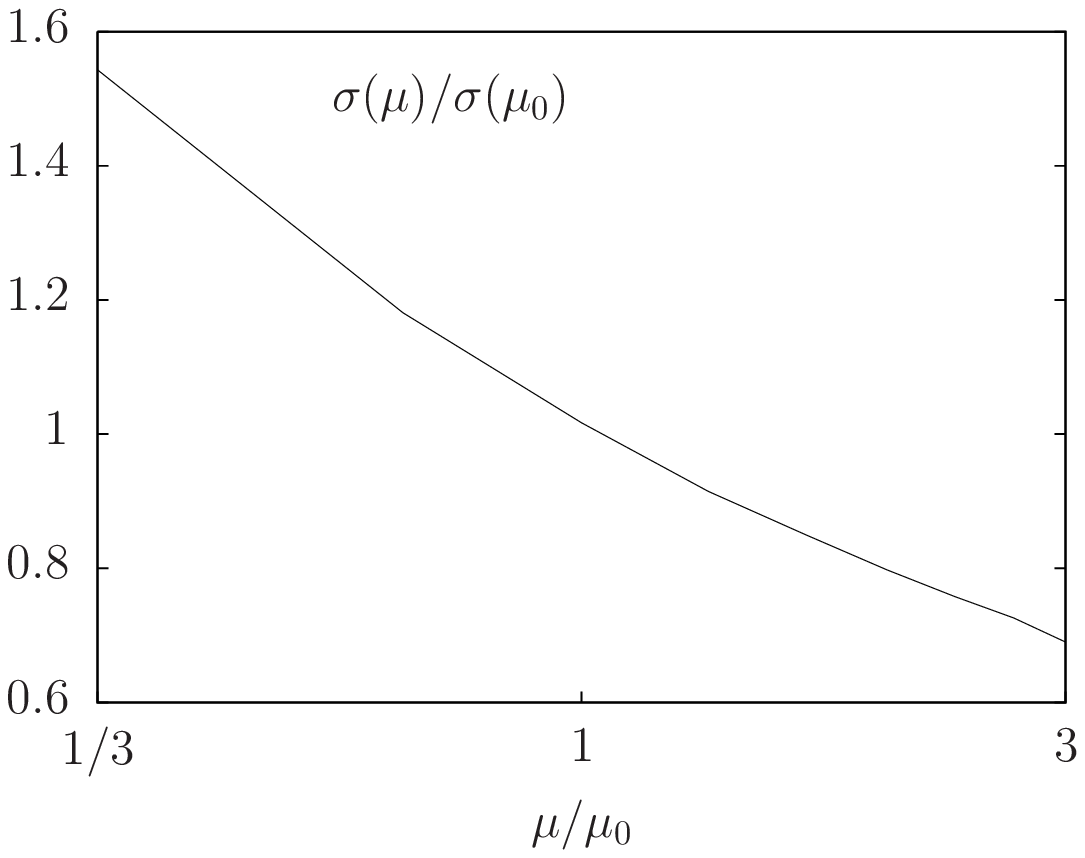}} 
\end{picture}
\caption{Renormalisation and factorisation scale dependence ($\mu=\mu_R=\mu_F$) of the 2-Higgs (left, $\mu_0=2m_H$) and 3-Higgs (right, $\mu_0=3m_H$) cross sections at the LHC.}
\label{fig:mu}
\end{figure}
Furthermore, the gluon fusion cross section is proportional to $\alpha_s^2$
and thus very sensitive to renormalisation scale variations,
as exemplified in Fig.~\ref{fig:mu} for $\mu = \mu_R=\mu_F$.  Here,
the scale $\mu$ is varied around the central choice $\mu_0 = 2\,m_H$ and
 $\mu_0 = 3\,m_H$ for 2- and 3-Higgs boson production, respectively, by a factor $\mu/\mu_0\in [1/3,3]$.  From this we estimate a scale uncertainty of about 50\%. 
Thus, large $K$-factors due to higher order effects can be expected.
 In Table~\ref{Tab:scale} we study the scale dependence of the cross section for 3-Higgs 
boson production
 by varying the renormalisation and 
 factorisation scales independently for $m_H=120\textrm{ GeV}$.  
 \begin{table}
\vspace*{0.8cm}
\centering
\begin{tabular}{cccc}
   \hline
   \hline
  $\sigma [10^{-2}{\rm fb}]$ & $\mu_F=m_H$ & $3 m_H$ & $9 m_H$ \\[0.5mm]
    \hline
   $\mu_R=m_H$  & $ 9.71 $ & $ 8.61$ & $ 7.66$ \\
   $3 m_H    $  & $ 7.21 $ & $ 6.38$ & $ 5.68 $ \\
   $9 m_H    $  & $ 5.57 $ & $ 4.93$ & $ 4.39 $ \\
  \hline
  \hline
\end{tabular}
\caption{\label{Tab:scale}Renormalisation and factorisation scale dependence of the $gg\to HHH$ cross section for $m_H=120 \;\textrm{GeV}$ at the LHC. }
\end{table}
Note that the gluon luminosity decreases with increasing scale $\mu_F$,
because in 3-Higgs boson production the momentum of the gluons has to be relatively high.
Hence, the cross section shrinks with increasing $\mu_F$ and $\mu_R$.
The table demonstrates that varying $\mu_R$ and $\mu_F$ in the same direction
yields a conservative estimate of the scale uncertainty.

\begin{figure}[htb]
\unitlength=1mm
\begin{picture}(150,60)
\put(0, 0){\includegraphics[width=7.cm]{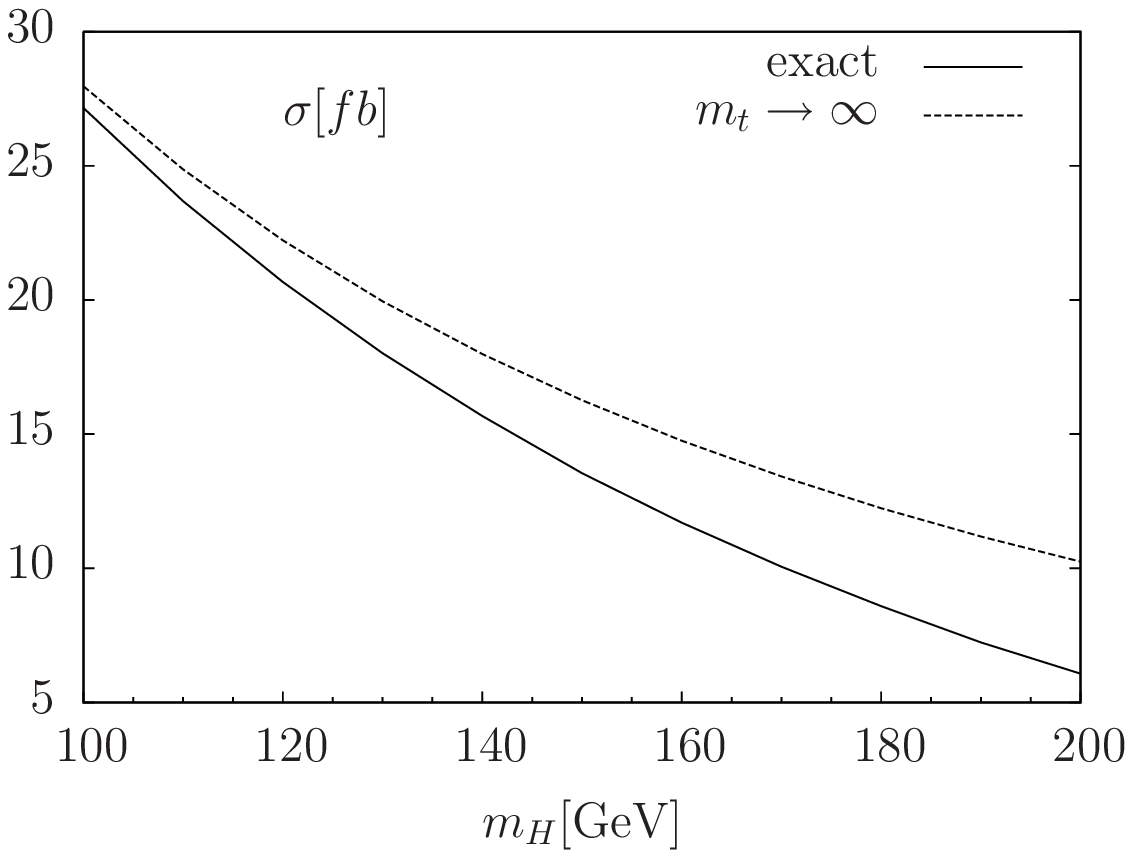}} 
\put(75,0){\includegraphics[width=7.cm]{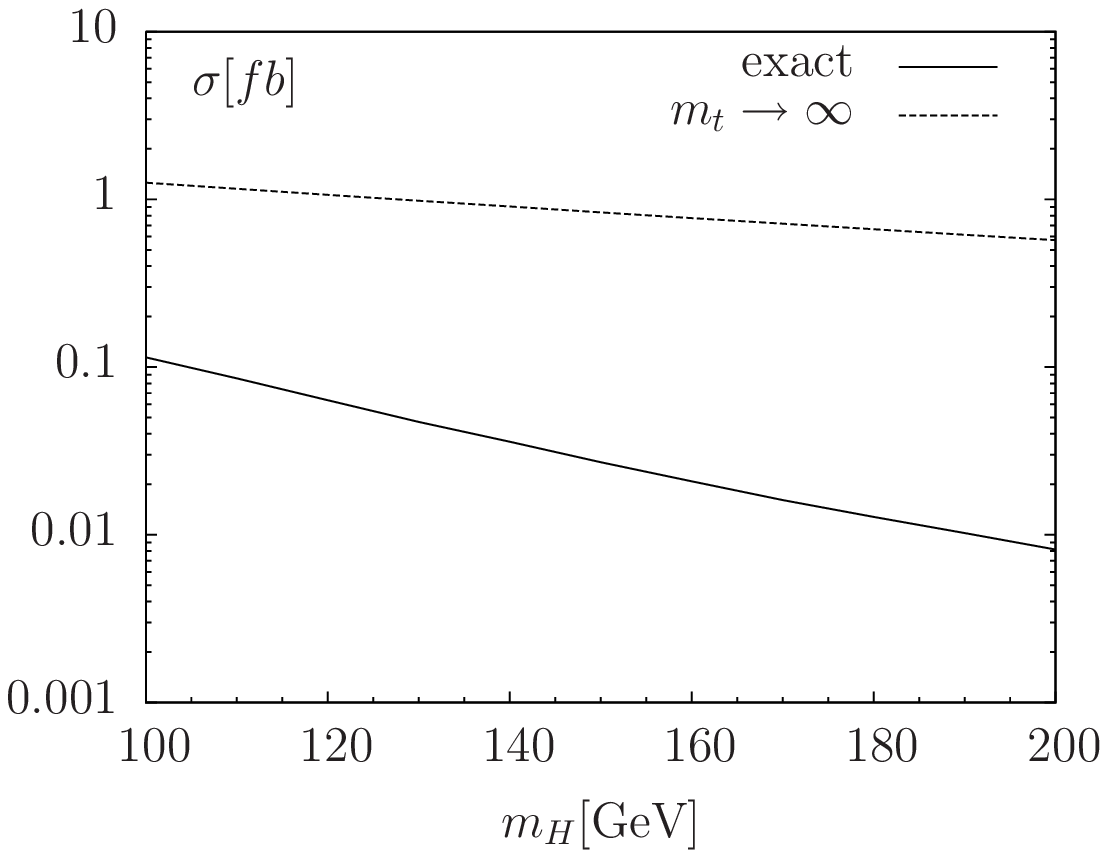}} 
\end{picture}
\caption{Comparison of the total cross section for 2-Higgs (left) and 3-Higgs (right) boson production vs.~$m_H$ at the LHC,
calculated for the physical value of $m_t$ and in the $m_t\to\infty$ limit.}
\label{fig:eff}
\end{figure}

\begin{figure}[htb]
\unitlength=1mm
\begin{picture}(150,40)
\put(0, 0){\includegraphics[width=7.cm]{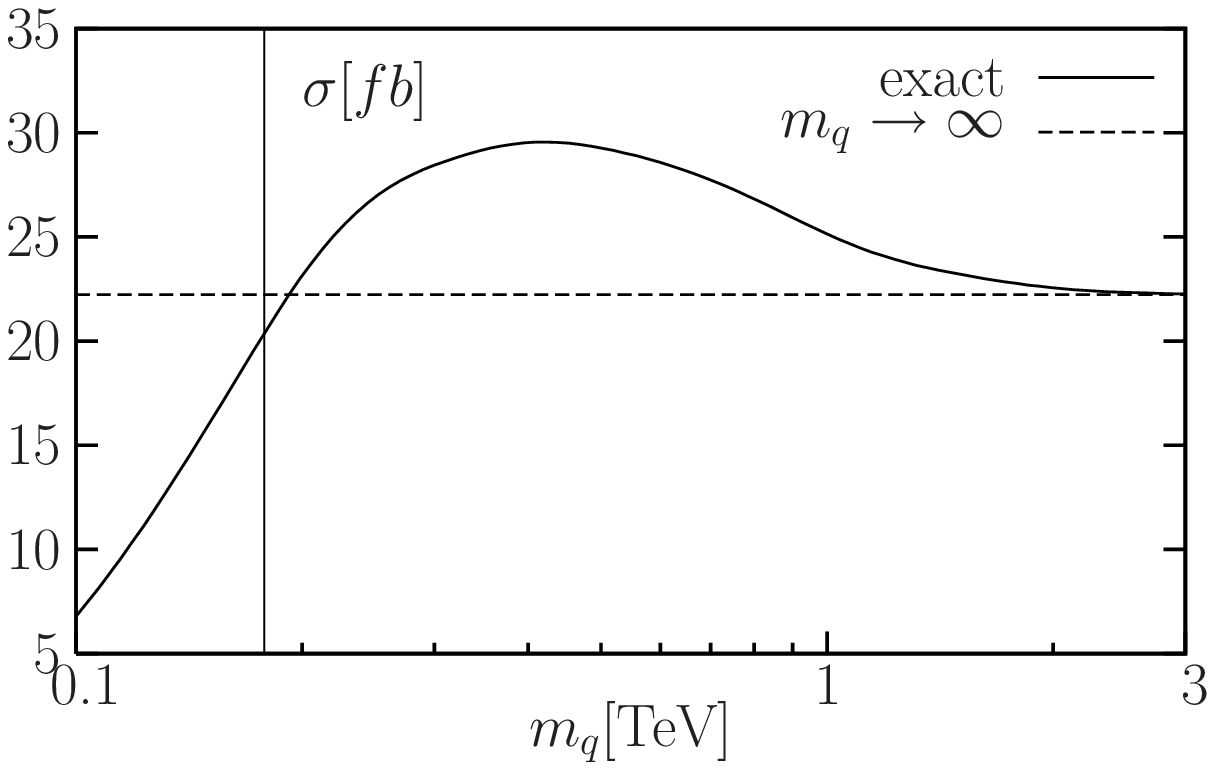}} 
\put(75,0){\includegraphics[width=7.cm]{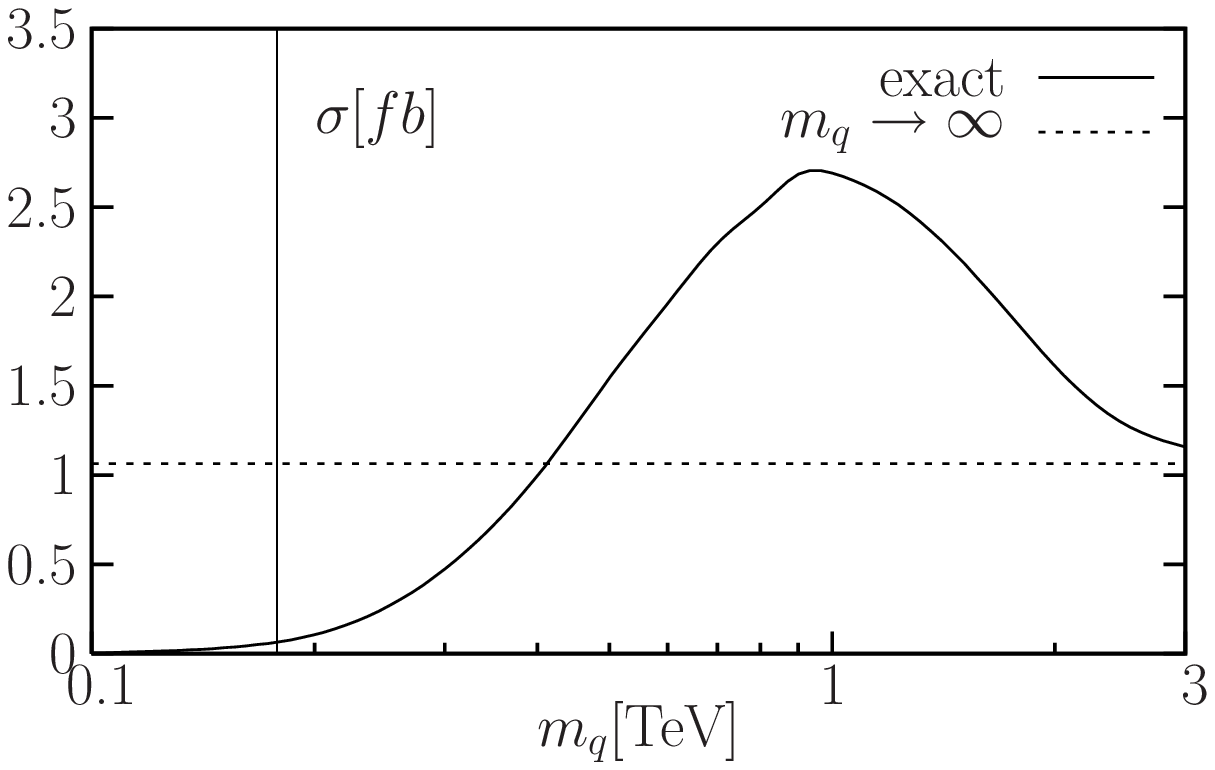}} 
\end{picture}
\caption{The total cross section for 2-Higgs (left) and 3-Higgs (right) production vs.~$m_q$ at the LHC.
The $m_q \to \infty$ limit is shown as horizontal line.
The vertical line indicates the actual value of the top quark mass.}
\label{fig:mttoinfty}
\end{figure}

Many years ago multi-Higgs boson production via gluon-fusion was studied in \cite{Glover:1988ky}
in the heavy top limit.
For single Higgs boson production via gluon fusion this limit is well known to be a
good approximation \cite{Spira:1995rr}.
In the context of multi-Higgs boson production,
the heavy top limit has been applied at the leading \cite{Glover:1988ky}
and next-to-leading level \cite{Dawson:1998py}.
In \cite{Baur:2002qd}, the quality of the heavy top  approximation has been 
studied for Higgs pair production, and  agreement at the
${\cal O}$(10\%)-level for the total cross section, but large discrepancies 
 for kinematic distributions have been observed when comparing results for
$m_t\to\infty$ and physical $m_t$.
In Fig.~\ref{fig:eff}
we compare our physical-$m_t$ results with results in the heavy top limit.
While there is reasonable agreement in the 2-Higgs case for small Higgs masses, 
the heavy top limit fails completely
in the 3-Higgs case. To better understand this observation, we study
 the variation of the cross sections
with the internal quark mass $m_q$ for a fixed value of $m_H=120$ GeV in Fig.~\ref{fig:mttoinfty}.
We see that in the 2-Higgs case the heavy top limit accidentally
agrees with the result for $m_q=m_t$ (indicated by the vertical
line).
However, asymptotically the result for finite $m_q$ approaches the $m_q\to \infty$
limit only for masses around 3 TeV.
The same holds for the 3-Higgs case, but here the result
for $m_q=m_t$ is an order of magnitude smaller than the heavy top
limit. 
In both cases the dominant contribution to the cross section comes
from the kinematic regime close to the top pair
threshold $s_{ij}\sim 4 m_t^2$. 
We conclude that the heavy top limit 
is not applicable when calculating multi-Higgs boson production 
cross sections and should not be used in experimental studies.

\subsection{Multi-Higgs boson production beyond the SM}

\begin{figure}[htb]
\unitlength=1mm
\begin{picture}(150,65)
\put(30,00){\includegraphics[width=11.cm]{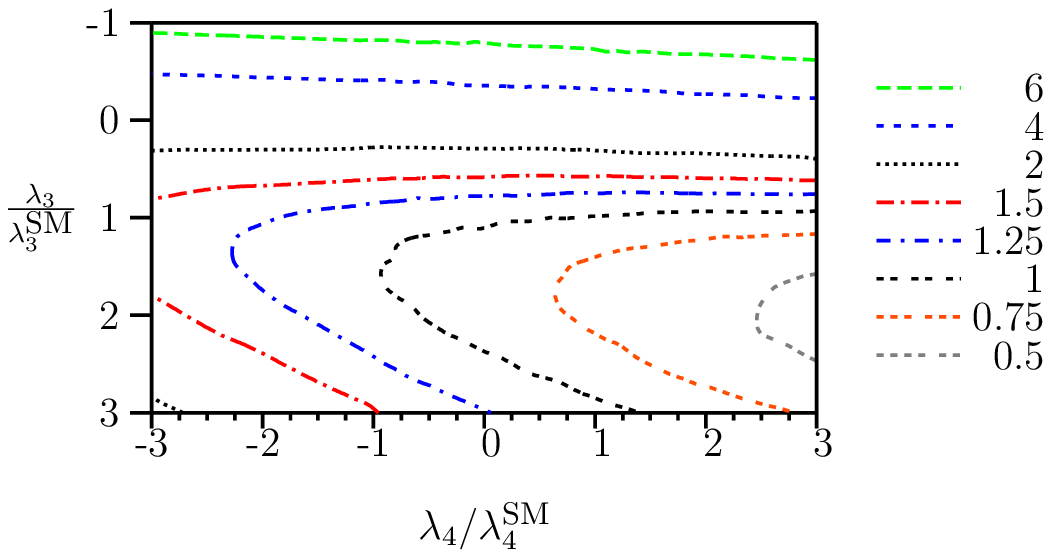}} 
\end{picture}
\caption{Contour plot showing  the variation of the cross section for 3-Higgs 
boson production
with $\lambda_3$ and $\lambda_4$  for $m_H=120$ GeV at the LHC. The numbers denote
the cross section normalised to 
$\sigma_{\textrm{SM}}(m_H=120\; \textrm{GeV})=0.0623\; \textrm{fb}$.}
\label{contour:l3l4}
\end{figure}

\begin{figure}[htb]
\unitlength=1mm
\begin{picture}(150,50)
\put(20,0){\includegraphics[width=10.cm]{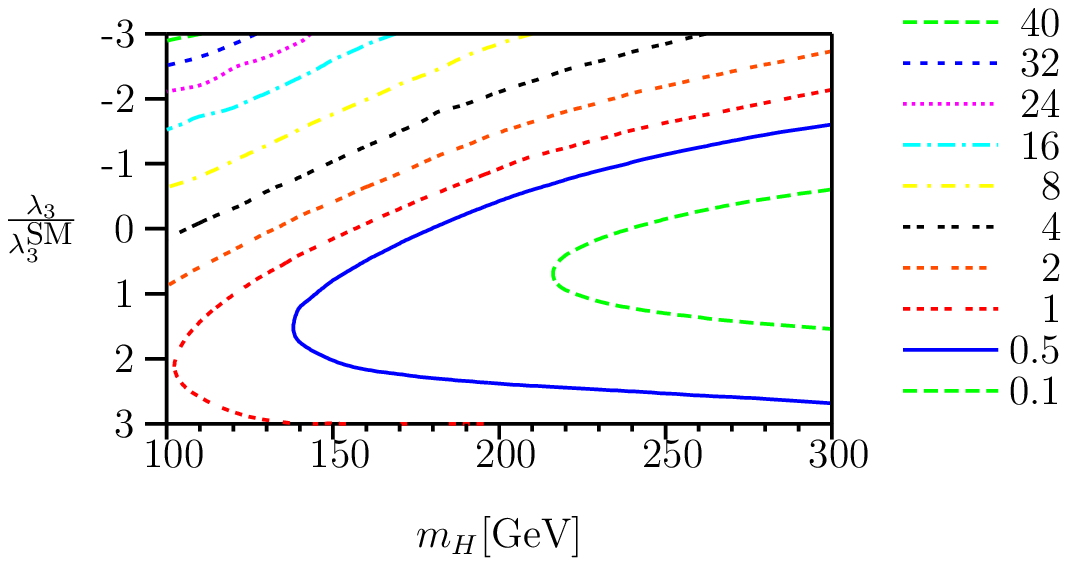}} 
\end{picture}
\caption{Contour plot showing the variation of the cross section for 3-Higgs 
boson production 
with $\lambda_3$ and $m_H$  for  $\lambda_4=\lambda_4^{SM}$ at the LHC, normalised to
$\sigma_{\textrm{SM}}(m_H=120\; \textrm{GeV})=0.0623\; \textrm{fb}$.}
\label{contour:l3mh}
\end{figure}

If one allows for higher dimensional operators in the Higgs sector
the trilinear and quartic Higgs self couplings are no longer directly 
related to the  Higgs mass. 
This has motivated us to study the variation of the 
cross section with respect to $\lambda_3$, $\lambda_4$ and $m_H$.
In Fig.~\ref{contour:l3l4} we illustrate
the variation of the cross section for 3-Higgs 
boson production
with $\lambda_3$ and $\lambda_4$ ($m_H$ is fixed), and in Fig.~\ref{contour:l3mh}
the variation with $\lambda_3$ and $m_H$ ($\lambda_4$ is fixed).

\begin{figure}[htb]
\vspace*{1.2cm}
\unitlength=1mm
\begin{picture}(150,50)
\put(30, 0){\includegraphics[width=7.cm]{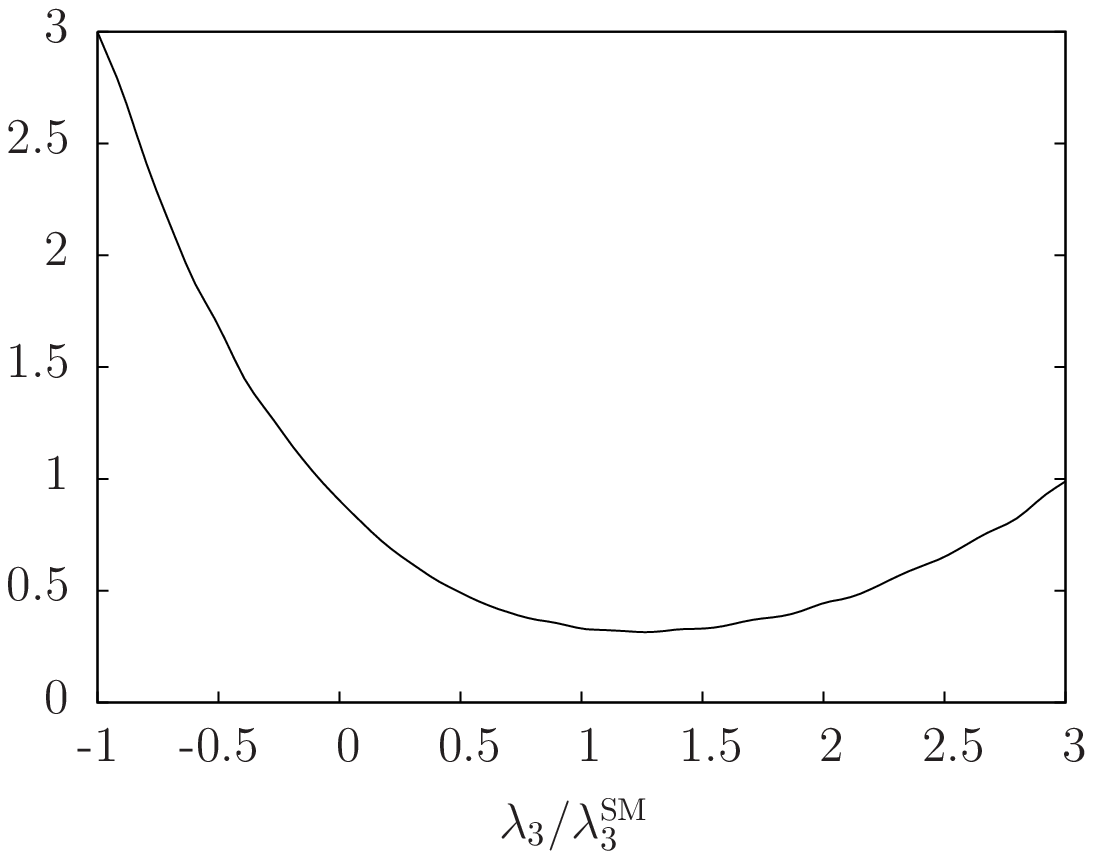}} 
\end{picture}
\caption{Variation of the cross section for 3-Higgs 
boson production with $\lambda_3$ for  
$m_H=160$ GeV and fixed  $\lambda_4=\lambda_4^{\textrm{SM}}$ at the LHC, normalised to
$\sigma_{\textrm{SM}}(m_H=120\; \textrm{GeV})=0.0623\; \textrm{fb}$.}
\label{fig:parabola}
\end{figure}

The variation of the cross section with
$\lambda_3$, $\lambda_4$ and $m_H$ is mainly due to phase space 
and interference effects. When $m_H$ increases, the phase space is reduced
and the PDF and $\alpha_s$ are to be taken at a higher value of $x$ and a
larger scale. 
All effects conspire and lead to a smaller cross section.  
The dependence on $\lambda_3$ is mainly due to the
 interference pattern. In Fig.~\ref{fig:parabola} a slice of the
 contour plot  Fig.~\ref{contour:l3mh} is shown for $m_H=160$ GeV.
When the trilinear coupling $\lambda_3$ becomes negative, the
destructive interference between box and pentagon topologies 
turns into a constructive one, which explains the rise of the
cross section in this regime. Increasing the coupling to
positive values beyond the SM value first enhances the
destructive interference effect, but finally the box topologies win and
the full cross section grows again. The minimum moves to lower values of $\lambda_3$
when the Higgs mass gets larger, due to the growth of the Higgs self-couplings.
We note that the same effect happens in the 2-Higgs case.
It implies that even a precise measurement of the cross section
alone would not lead to a unique determination of the trilinear coupling. 

This reasoning allows now to understand the variation of the cross section with 
$\lambda_4$ in Fig.~\ref{contour:l3l4}. For negative values
of $\lambda_3$ the box and pentagon topologies interfere constructively
and the relative importance of the triangle contribution proportional 
to $\lambda_4,\lambda_3^2$ is further reduced, 
resulting in an almost flat dependence on $\lambda_4$.
For positive $\lambda_3$ the 
contribution of the triangle topologies
is pronounced by destructive
interferences  between the various
topologies leading to a slight variation
with $\lambda_4$.  As can be seen from Fig.~\ref{contour:l3l4},
for $\lambda_3/\lambda_3^\text{SM}$
in the range 0.5 to 1.5 and $\lambda_4/\lambda_4^\text{SM}$ in the range -3 to 3
the cross section varies from 0.03 to 0.1 fb.

In principle the couplings are restricted only by unitarity bounds.
To illustrate the effects of large couplings
that approach the non-perturbative regime,
we list in Table~\ref{Tab:extreme} cross sections for values  
$|\lambda_3/v|, |\lambda_4| \sim 4\pi$. We see that non-perturbative effects in the Higgs sector may well lead to sizable
triple Higgs cross sections of up to 30 fb at the LHC.
\begin{table} 
\centering
\begin{tabular}{cccc}
   \hline
   \hline
  $\sigma [\textrm{fb}]$ & $\lambda_4=-4\pi$ & $0$ & $4\pi$ \\[0.5mm]
    \hline
  $\lambda_3/v=-4\pi$  & $ 28.0 $ & $ 30.7$ & $  33.4$ \\
            $0    $  & $ 0.169 $ & $ 0.0271$ & $ 0.0428 $ \\
            $ 4\pi$  & $ 12.2 $ & $ 14.0$ & $ 15.8 $ \\
  \hline
  \hline
\end{tabular}

\caption{\label{Tab:extreme}3-Higgs 
boson production cross sections for extreme choices of the Higgs couplings for $m_H=150$ GeV at the LHC.}
\end{table}

\begin{figure}[htb]
\vspace*{1.2cm}
\unitlength=1mm
\begin{picture}(150,50)
\put(-3,48){a)}
\put(0, 0){\includegraphics[width=7.cm]{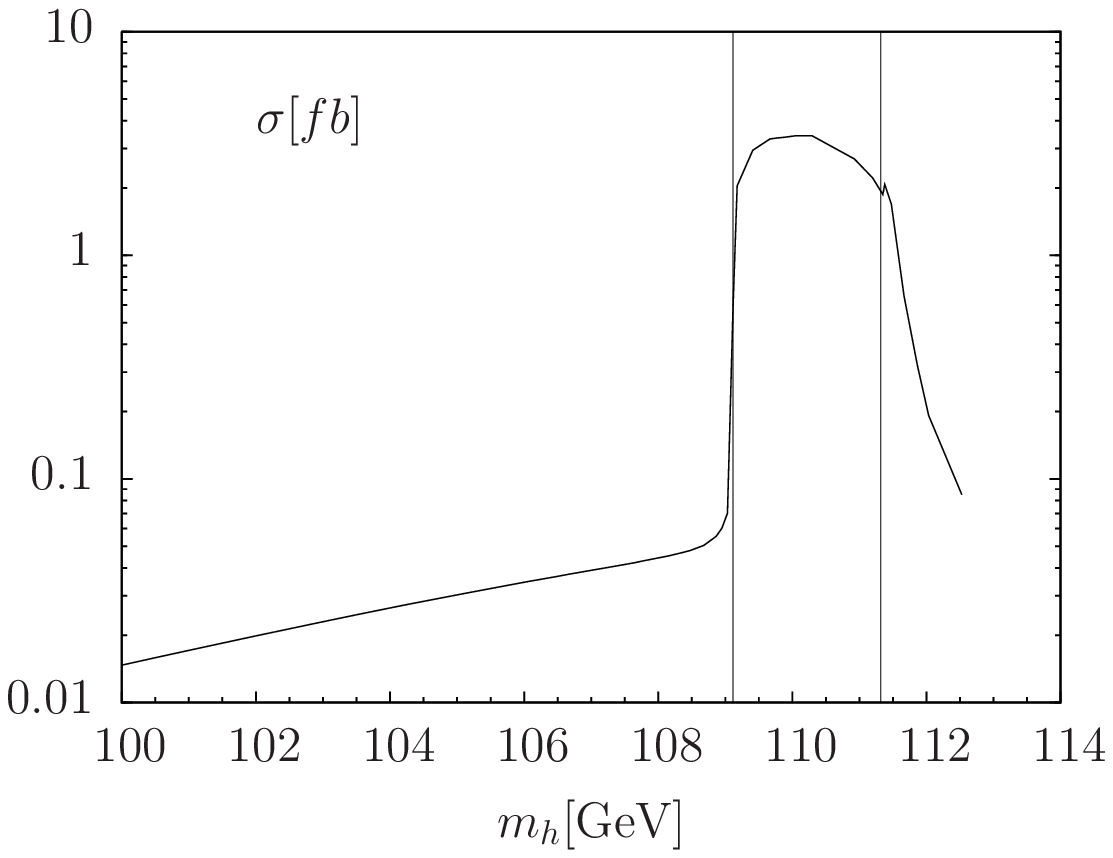}} 
\put(72,48){b)}
\put(75,0){\includegraphics[width=7.cm]{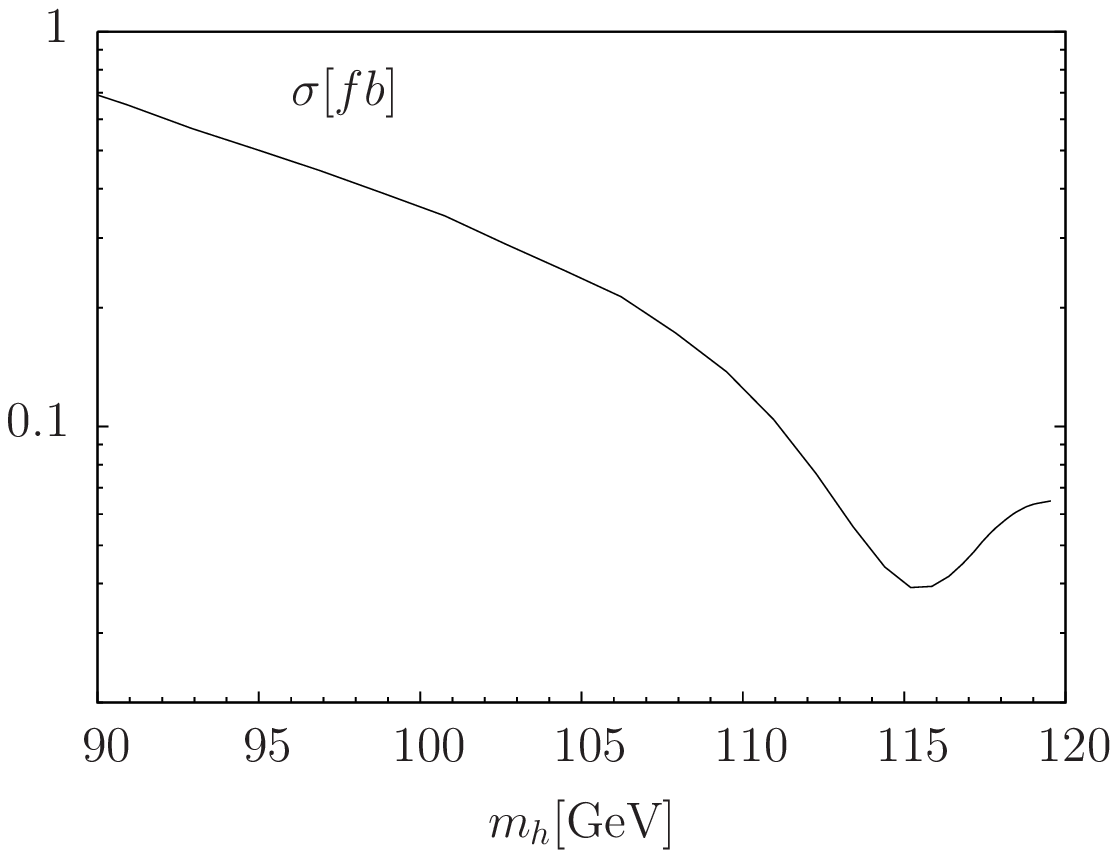}} 
\end{picture}
\caption{\label{fig:susy}Cross section for 3-Higgs 
boson production vs.~$m_h$ at the LHC in the MSSM for $\tan\beta=3$ (left)
and $\tan\beta=50$ (right) including mixing effects ($A_t=1 \textrm{TeV}$,
$\mu/\textrm{TeV}=-1\,(+1)$ for $\tan\beta=3\,(50)$).
The vertical lines in the left plot
indicate the 2- and 3-Higgs boson thresholds ($H\to hh$ and $H\to hhh$).}
\end{figure}

As already pointed out in Section 2, in 2HDMs one finds two
amplification effects for multi-Higgs boson production. 
Both effects are illustrated in Fig.~\ref{fig:susy} for the MSSM.
For small $\tan\beta=3$ the heavy CP even  Higgs boson may become
resonant in some of the topologies sketched in Fig.~\ref{graphs}. This is illustrated in Fig.~\ref{fig:susy}a.
With the given choice of parameters one sees that for $m_h>109$ GeV
the $H\to hh$ channel opens up inducing a resonant amplification
of the box and triangle topologies, $B$ and $T_2$. For $m_h>111$ GeV
also the $H\to hhh$ channel opens up leading to an enhancement 
of the triangle topology $T_1$ proportional to the quartic 
coupling $\lambda_{Hhhh}$. Since this triangle contribution
is suppressed relative to the other topologies the effect is hardly
visible in the plot (tiny peak next to the right vertical line).
We see that in BSM scenarios resonant amplification may lead to
triple Higgs production reaching a few
fb (not taking into account the expected $K$-factor of 2),
which would be observable at the SuperLHC or even at the LHC.

For large $\tan\beta=50$  the top contribution is largely
suppressed and the bottom loops become dominant due to the enhanced 
Yukawa couplings.
Resonance effects on the other hand are now negligible, because the Higgs 
width $\Gamma_H$ is a factor 50 larger than at small $\tan \beta=3$.
Based on the double Higgs production  results in \cite{Jin:2005gw}, we expect
therefore that the LO triple Higgs production cross section in bottom quark
fusion is smaller than in gluon fusion.
As the pentagon contribution to the cross section is proportional
to $\lambda_{hb\bar b}^6$, all other topologies are much less relevant.
For $m_A\to \infty$, $m_h$ approaches its maximal value.  
In this limit one  obtains
the SM value for the cross section as shown in Fig.~\ref{fig:susy}b. 
The minimum around $m_h\sim 115$ GeV arises due to the interplay of
the Yukawa- and Higgs self coupling.
Although the cross section rises for
decreasing $m_h$, it is always well below 1 fb in the range considered.
Note that in  the 2-Higgs boson production case the bottom loops are more
pronounced than in the 3-Higgs case, because
the scalar loop integrals are probed in different kinematical regions.
  
In principle, also squark loops have to be considered
to obtain a complete prediction for $hhh$-production in 
the MSSM. In the heavy squark limit this contribution 
decouples and approaches zero in contrast to the quark contribution.
For the present illustration of amplification 
effects we have neglected squark effects.

\clearpage

\section{Summary}

We presented our calculation of the loop-induced 
processes $gg\to HH$ and $gg\to HHH$, and discussed the resulting
cross sections and their experimental accessibility at the LHC in
the SM and beyond.
The contributions from
pentagon, box and triangle topologies exhibit strong interference patterns
and large differences in the equal and opposite gluon helicity components.
Generally, the opposite helicity component, which
corresponds to a gluon pair with helicity 2, is suppressed by more than an
order of magnitude. Furthermore, the triangle topologies are suppressed relative
to the box and pentagon topologies. This results in a complex
dependence of the  cross sections on the trilinear and quartic Higgs self couplings.
The cross section for 3-Higgs boson production varies strongly with
the quartic Higgs self coupling.  For $m_H=100$ GeV and $200$ GeV
it affects the cross section by +1\%  and $-57.5$\%, respectively.
This effect is due to the fact that the Higgs self coupling contributions
are enhanced for higher Higgs masses and that there is a strong 
destructive interference between box and pentagon contributions.
The estimated renormalisation and factorisation
scale uncertainty of about 50\% (variation by a factor 3) is typical
for leading-order QCD cross sections.
We also investigated the applicability of the heavy top quark
approximation for 2- and 3-Higgs boson production and
demonstrated that the heavy top limit 
is not applicable when calculating multi-Higgs boson production 
cross sections.
In summary, we find that the SM cross sections for 3-Higgs boson production
are too small for observation at the LHC.  The measurement of the SM trilinear 
Higgs self coupling in 2-Higgs boson production, on the other hand, may be feasible at a luminosity upgraded LHC, 
termed SuperLHC, collecting 6000 fb$^{-1}$ of data,
as long as QCD backgrounds are not prohibitively large \cite{Weiglein:2004hn,Gianotti:2002xx}.

The experimental prospects improve if favourable extensions of the SM
are realized in nature.
We have demonstrated that cross sections for
triple Higgs boson production can be as large as 
$\mathcal{O}(10)$ fb if one allows for higher dimensional 
operators or considers 2-Higgs-doublet models, e.g. the MSSM. 
Here, two amplification effects
have been analysed. First, we studied Yukawa coupling enhancements 
through mixing that lead to important bottom loop contributions.
For the 3-Higgs case the cross section remains below 1 fb for the
parameters considered and $\tan\beta=50$. Secondly, we demonstrated
that resonance effects due to an internal heavy Higgs boson decaying into
2 or 3 light Higgs bosons can enhance the 3-Higgs cross section
into the potentially observable $\mathcal{O}$(fb) region.
Higher dimensional operators lead to   
essentially unconstrained trilinear and quartic couplings. 
Strong coupling of the order 
$|\lambda_3/v|, |\lambda_4| \sim 4\pi$ leads to   
cross sections of up to 30 fb.
This implies that already the LHC should be able to restrict
the $\lambda_3$-$\lambda_4$ plane. Although these bounds will not be very restrictive,
they may still exclude parameter regions that exhibit non-perturbative effects,
which would be an important qualitative finding.
A more stringent bound on $\lambda_3$ alone will
be obtained from Higgs pair production.
We conclude that multi-Higgs boson production at the luminosity-upgraded LHC
is an interesting probe of Higgs sectors beyond the SM and warrants further study.
A 200 TeV VLHC would of course further improve the 
sensitivity.

\section*{Acknowledgements}

S.K.~thanks the PPT group at the University of Edinburgh
for their hospitality while this paper was completed. 
T.B.~thanks F.~Boudjema for discussion.
This work  was supported by 
the Deutsche Forschungsgemeinschaft (DFG) under contract number BI 1050/1, the
DFG Research Training Group 1147,
the Scottish Universities Physics Alliance (SUPA),
and the Bundesministerium f\"ur Bildung und Forschung (BMBF, Bonn, Germany)
under contract number 05HT1WWA2.


\begin{thebibliography}{99}


\bibitem{Higgs:1964ia}
  P.~W.~Higgs,
  Phys.\ Lett.\  {\bf 12} (1964) 132;
  P.~W.~Higgs,
  Phys.\ Rev.\  {\bf 145} (1966) 1156;
  F.~Englert and R.~Brout,
  Phys.\ Rev.\ Lett.\  {\bf 13} (1964) 321;
  G.~S.~Guralnik, C.~R.~Hagen and T.~W.~B.~Kibble,
  Phys.\ Rev.\ Lett.\  {\bf 13} (1964) 585.

\bibitem{Buscher:2005re}
  V.~Buscher and K.~Jakobs,
  %
  Int.\ J.\ Mod.\ Phys.\ A {\bf 20} (2005) 2523
  [arXiv:hep-ph/0504099].
  
\bibitem{Djouadi:2005gi}
  A.~Djouadi,
  %
  arXiv:hep-ph/0503172 and
  arXiv:hep-ph/0503173.
\bibitem{Binoth:1998tg}
  T.~Binoth and A.~Ghinculov,
  Nucl.\ Phys.\ B {\bf 550} (1999) 77 
  [arXiv:hep-ph/9808393].

\bibitem{Ghinculov:1998km}
  A.~Ghinculov, T.~Binoth and J.~J.~van der Bij,
  Phys.\ Lett.\ B {\bf 427} (1998) 343
  [arXiv:hep-ph/9802367].



\bibitem{Barate:2003sz}
  R.~Barate {\it et al.}  [LEP Working Group for Higgs boson searches],
  Phys.\ Lett.\ B {\bf 565} (2003) 61
  [arXiv:hep-ex/0306033].

\bibitem{Duhrssen:2004cv}
  M.~D\"uhrssen, S.~Heinemeyer, H.~Logan, D.~Rainwater, G.~Weiglein and D.~Zeppenfeld,
  Phys.\ Rev.\ D {\bf 70} (2004) 113009
  [arXiv:hep-ph/0406323].

\bibitem{Weiglein:2004hn}
  G.~Weiglein {\it et al.}  [LHC/LC Study Group],
   ``Physics interplay of the LHC and the ILC,''
  %
  arXiv:hep-ph/0410364.



  
\bibitem{Djouadi:1999gv}
  A.~Djouadi, W.~Kilian, M.~Muhlleitner and P.~M.~Zerwas,
  %
  Eur.\ Phys.\ J.\ C {\bf 10} (1999) 27
  [arXiv:hep-ph/9903229].
  
  
\bibitem{Battaglia:2001nn}
  M.~Battaglia, E.~Boos and W.~M.~Yao,
  %
in {\it Proc. of the APS/DPF/DPB Summer Study on the Future of Particle Physics (Snowmass 2001) } ed. N.~Graf,
  eConf {\bf C010630} (2001) E3016
  [arXiv:hep-ph/0111276].

\bibitem{Gutierrez-Rodriguez:2006qk}
  A.~Gutierrez-Rodriguez, M.~A.~Hernandez-Ruiz and O.~A.~Sampayo,
  %
  arXiv:hep-ph/0601238.

\bibitem{Jikia:1992mt}
  G.~V.~Jikia,
  Nucl.\ Phys.\ B {\bf 412} (1994) 57.

\bibitem{Boudjema:1995cb}
  F.~Boudjema and E.~Chopin,
  Z.\ Phys.\ C {\bf 73} (1996) 85
  [arXiv:hep-ph/9507396].


\bibitem{Belusevic:2004pz}
  R.~Belusevic and G.~Jikia,
  Phys.\ Rev.\ D {\bf 70} (2004) 073017
  [arXiv:hep-ph/0403303].

  
\bibitem{Djouadi:1999rc}
  A.~Djouadi, W.~Kilian, M.~Muhlleitner and P.~M.~Zerwas,
  %
  Eur.\ Phys.\ J.\ C {\bf 10} (1999) 45
  [arXiv:hep-ph/9904287].

\bibitem{Baur:2002qd}
  U.~Baur, T.~Plehn and D.~L.~Rainwater,
  %
  Phys.\ Rev.\ D {\bf 67} (2003) 033003
  [arXiv:hep-ph/0211224].


\bibitem{Baur:2002rb}
  U.~Baur, T.~Plehn and D.~L.~Rainwater,
  %
  Phys.\ Rev.\ Lett.\  {\bf 89} (2002) 151801
  [arXiv:hep-ph/0206024].

\bibitem{Baur:2003gp}
  U.~Baur, T.~Plehn and D.~L.~Rainwater,
  %
  Phys.\ Rev.\ D {\bf 68} (2003) 033001
  [arXiv:hep-ph/0304015].


\bibitem{Baur:2003gpx}
  U.~Baur, T.~Plehn and D.~L.~Rainwater,
  %
  Phys.\ Rev.\ D {\bf 69} (2004) 053004
  [arXiv:hep-ph/0310056].

\bibitem{Dawson:2006dm}
  S.~Dawson, C.~Kao, Y.~Wang and P.~Williams,
  arXiv:hep-ph/0610284.
  
\bibitem{Jin:2005gw}
  L.~G.~Jin, C.~S.~Li, Q.~Li, J.~J.~Liu and R.~J.~Oakes,
  Phys.\ Rev.\ D {\bf 71} (2005) 095004
  [arXiv:hep-ph/0501279].

\bibitem{Glover:1987nx}
  E.~W.~N.~Glover and J.~J.~van der Bij,
  Nucl.\ Phys.\ B {\bf 309} (1988) 282.

\bibitem{Plehn:1996wb}
  T.~Plehn, M.~Spira and P.~M.~Zerwas,
  Nucl.\ Phys.\ B {\bf 479} (1996) 46
  [Erratum-ibid.\ B {\bf 531} (1998) 655]
  [arXiv:hep-ph/9603205].

\bibitem{Krause:1997rc}
  A.~Krause, T.~Plehn, M.~Spira and P.~M.~Zerwas,
  Nucl.\ Phys.\ B {\bf 519} (1998) 85
  [arXiv:hep-ph/9707430].

\bibitem{Brein:1999sy}
  O.~Brein and W.~Hollik,
  Eur.\ Phys.\ J.\ C {\bf 13} (2000) 175
  [arXiv:hep-ph/9908529].

\bibitem{Plehn:2005nk}
  T.~Plehn and M.~Rauch,
  %
  Phys.\ Rev.\ D {\bf 72} (2005) 053008
  [arXiv:hep-ph/0507321].
\bibitem{Kanemura:2002vm}
  S.~Kanemura, S.~Kiyoura, Y.~Okada, E.~Senaha and C.~P.~Yuan,
  %
  Phys.\ Lett.\ B {\bf 558} (2003) 157
  [arXiv:hep-ph/0211308].


\bibitem{Kanemura:2004mg}
  S.~Kanemura, Y.~Okada, E.~Senaha and C.~P.~Yuan,
  %
  Phys.\ Rev.\ D {\bf 70} (2004) 115002
  [arXiv:hep-ph/0408364].


\bibitem{Appelquist:1974tg}
  T.~Appelquist and J.~Carazzone,
  Phys.\ Rev.\ D {\bf 11} (1975) 2856.

\bibitem{Barger:2003rs}
  V.~Barger, T.~Han, P.~Langacker, B.~McElrath and P.~Zerwas,
  %
  Phys.\ Rev.\ D {\bf 67} (2003) 115001
  [arXiv:hep-ph/0301097].



\bibitem{Dib:2005re}
  C.~O.~Dib, R.~Rosenfeld and A.~Zerwekh,
  JHEP {\bf 0605} (2006) 074
  [arXiv:hep-ph/0509179].



\bibitem{Lee:1977eg}
  B.~W.~Lee, C.~Quigg and H.~B.~Thacker,
   ``Weak Interactions At Very High-Energies: The Role Of The Higgs Boson
  Mass,''
  Phys.\ Rev.\ D {\bf 16} (1977) 1519.

\bibitem{Binoth:1996au}
  T.~Binoth and J.~J.~van der Bij,
  %
  Z.\ Phys.\ C {\bf 75} (1997) 17
  [arXiv:hep-ph/9608245].

\bibitem{Gunion:1989we}
  J.~F.~Gunion, H.~E.~Haber, G.~L.~Kane and S.~Dawson,
  ``The Higgs Hunter's Guide,''
SCIPP-89/13;
   ``Errata for the Higgs hunter's guide,''
  %
  arXiv:hep-ph/9302272.


\bibitem{Haber:1993an}
  H.~E.~Haber and R.~Hempfling,
  Phys.\ Rev.\ D {\bf 48} (1993) 4280
  [arXiv:hep-ph/9307201].

\bibitem{Carena:1995bx}
  M.~Carena, J.~R.~Espinosa, M.~Quiros and C.~E.~M.~Wagner,
   ``Analytical expressions for radiatively corrected Higgs masses and couplings
  in the MSSM,''
  Phys.\ Lett.\ B {\bf 355} (1995) 209
  [arXiv:hep-ph/9504316].

\bibitem{HDECAY}
A.~Djouadi, J.~Kalinowski and M.~Spira, Comput.\ Phys.\ Commun.\ 108 (1998) 56.
%

\bibitem{Boudjema:2001ii}
  F.~Boudjema and A.~Semenov,
  Phys.\ Rev.\ D {\bf 66} (2002) 095007
  [arXiv:hep-ph/0201219].

\bibitem{Binoth:1999sp}
  T.~Binoth, J.~P.~Guillet and G.~Heinrich,
  Nucl.\ Phys.\ B {\bf 572} (2000) 361 [arXiv:hep-ph/9911342].

\bibitem{Binoth:2003xk}  
  T.~Binoth, J.~P.~Guillet and F.~Mahmoudi,
  JHEP {\bf 0402} (2004) 057 [arXiv:hep-ph/0312334].

\bibitem{Binoth:2005ua}
  T.~Binoth, M.~Ciccolini, N.~Kauer and M.~Kr\"amer,
  JHEP {\bf 0503} (2005) 065
  [arXiv:hep-ph/0503094].



\bibitem{Binoth:2005ff}
  T.~Binoth, J.~P.~Guillet, G.~Heinrich, E.~Pilon and C.~Schubert,
  JHEP {\bf 0510} (2005) 015
  [arXiv:hep-ph/0504267].

\bibitem{Binoth:2006rc}
  T.~Binoth, M.~Ciccolini and G.~Heinrich,
  arXiv:hep-ph/0601254.


\bibitem{Binoth:2006mf}
  T.~Binoth, A.~Guffanti, J.~P.~Guillet, S.~Karg, N.~Kauer and T.~Reiter,
  arXiv:hep-ph/0606318.

\bibitem{Xu:1986xb}
  Z.~Xu, D.~H.~Zhang and L.~Chang,
  %
  Nucl.\ Phys.\ B {\bf 291} (1987) 392.


\bibitem{Dixon:1995xx}
  L.~J.~Dixon,
  arXiv:hep-ph/9507214.


\bibitem{Nogueira:1991ex}
  P.~Nogueira,
  J.\ Comput.\ Phys.\  {\bf 105} (1993) 279.


\bibitem{Vermaseren:2000nd}
  J.~A.~M.~Vermaseren,
  arXiv:math-ph/0010025.

\bibitem{Hahn:2000kx} 
  T.~Hahn,
  Comput.\ Phys.\ Commun.\ {\bf 140} (2001) 418 [hep-ph/0012260].

\bibitem{Martin:2002aw}
  A.~D.~Martin, R.~G.~Roberts, W.~J.~Stirling and R.~S.~Thorne,
  Eur.\ Phys.\ J.\ C {\bf 28} (2003) 455
  [arXiv:hep-ph/0211080].


\bibitem{LHAPDF} http://hepforge.cedar.ac.uk/lhapdf/

\bibitem{Berends:1994pv}
  F.~A.~Berends, R.~Pittau and R.~Kleiss,
  Nucl.\ Phys.\ B {\bf 424} (1994) 308
  [arXiv:hep-ph/9404313].

\bibitem{Kleiss:1994qy}
  R.~Kleiss and R.~Pittau,
  Comput.\ Phys.\ Commun.\  {\bf 83} (1994) 141
  [arXiv:hep-ph/9405257].

\bibitem{Kauer:2001sp}
  N.~Kauer and D.~Zeppenfeld,
  Phys.\ Rev.\ D {\bf 65} (2002) 014021
  [arXiv:hep-ph/0107181].

\bibitem{Kauer:2002sn}
  N.~Kauer,
  Phys.\ Rev.\ D {\bf 67} (2003) 054013
  [arXiv:hep-ph/0212091].

\bibitem{BASES} http://minami-home.kek.jp/

\bibitem{FeynHiggs}
S.~Heinemeyer, W.~Hollik and G.~Weiglein, Comp. Phys. Commun. 124 (2000) 76; 
M.~Frank, S.~Heinemeyer, W.~Hollik and G.~Weiglein, arXiv:hep-ph/0202166.
%

\bibitem{Dawson:1990zj}
  S.~Dawson,
  Nucl.\ Phys.\ B {\bf 359} (1991) 283.

\bibitem{Djouadi:1991tk}
  A.~Djouadi, M.~Spira and P.~M.~Zerwas,
  Phys.\ Lett.\ B {\bf 264} (1991) 440.

\bibitem{Glover:1988ky}
  E.~W.~N.~Glover and J.~J.~van der Bij,
CERN-TH-5022-88
{\it Presented at 23rd Rencontres de Moriond: Current Issues in Hadron Physics, Les Arcs, France, Mar 13-19, 1988}.


\bibitem{Spira:1995rr}
  M.~Spira, A.~Djouadi, D.~Graudenz and P.~M.~Zerwas,
  Nucl.\ Phys.\ B {\bf 453} (1995) 17
  [arXiv:hep-ph/9504378].

\bibitem{Dawson:1998py}
  S.~Dawson, S.~Dittmaier and M.~Spira,
  Phys.\ Rev.\ D {\bf 58} (1998) 115012
  [arXiv:hep-ph/9805244].



\bibitem{Gianotti:2002xx}
  F.~Gianotti {\it et al.},
   ``Physics potential and experimental challenges of the LHC luminosity
  upgrade,''
  Eur.\ Phys.\ J.\ C {\bf 39} (2005) 293
  [arXiv:hep-ph/0204087].

\end{thebibliography}
\end{document}